\documentclass[12pt,preprint]{aastex}

\slugcomment{ApJ in press (Vol.569, April 20, 2002 issue)}

\shorttitle{\ion{Mg}{2} DLA APM 08279+5255}
\shortauthors{Kobayashi et al.}

\begin{document}


\title{\ion{Mg}{2} Absorption Lines in z=2.974 Damped Lyman-$\alpha$
       system\\ toward Gravitationally Lensed QSO APM 08279+5255:\\
       Detection of Small-scale Structure in \ion{Mg}{2}
       Absorbing Clouds\altaffilmark{1}}


\author{Naoto Kobayashi, Hiroshi Terada, Miwa Goto}
\affil{Subaru Telescope, National Astronomical Observatory Japan, 
       650 North A'ohoku Place, Hilo, HI 96720}
\email{naoto@naoj.org}

\and

\author{A. T. Tokunaga}
\affil{Institute for Astronomy, University of Hawaii,
       2680 Woodlawn Drive, Honolulu, HI 96822}


\altaffiltext{1}{Based on data collected at Subaru Telescope, which is
operated by the National Astronomical Observatory of Japan. }


\begin{abstract}
1.02$-$1.16\,$\mu$m spectra ($\lambda/\Delta\lambda\sim$7,000) of the
gravitationally lensed QSO APM 08279+5255 at z$_{em}$=3.911 were
obtained during the commissioning run of IRCS, the 1$-$5\,$\micron$
near-infrared camera and spectrograph for the Subaru 8.2 m
Telescope.
Strong \ion{Mg}{2} doublet $\lambda\lambda$2976,2800 and \ion{Fe}{2}
$\lambda$2600, \ion{Fe}{2} $\lambda$2587 absorption lines at
z$_{abs}$=2.974 were clearly detected in the rest-frame UV spectra,
confirming the presence of a damped Lyman-$\alpha$ system at the
redshift as suggested by Petitjean et al.  Also \ion{Mg}{1}
$\lambda$2853 absorption line is probably detected.  An analysis of the
absorption lines including velocity decomposition was performed.  This
is a first detailed study of \ion{Mg}{2} absorption system at high
redshift (z $>$ 2.5) where the \ion{Mg}{2} doublet shifts into the
near-infrared in the observer's frame.

The spectra of the lensed QSO pair A and B with 0\farcs38 separation
were resolved in some exposure frames under excellent seeing
condition. We extracted the \ion{Mg}{2} doublet spectra of A and B
separately.  Although three velocity components (v$\sim$-28, +5, +45 km
s$^{-1}$) are known to exist in this \ion{Mg}{2} system (Petitjean et
al.), the v$\sim$+45 km s$^{-1}$ absorption line was {\it not} detected
toward source B, showing that the +45 km s$^{-1}$ \ion{Mg}{2} cloud lies
only in the line of sight to the source A.  Our results suggests that
the size of the \ion{Mg}{2} absorbing clouds is as small as 200 pc,
which corresponds to the separation of A and B at the redshift of the
absorber.  This is the first direct detection of the small-scale
structure of \ion{Mg}{2} clouds at high-redshift, confirming the
estimated cloud sizes from photoionization model by Churchill and
Charlton.

\end{abstract}




\section{Introduction}

The gravitationally lensed QSO APM 08279+5255 (z$_{em}$=3.911) has been
given much attention since its discovery by Irwin et al. (1998). There
are two major gravitationally lensed images ``A'' and ``B'' with a
separation of 0\farcs38 and flux ratio of f$_B$/f$_A$=0.77 (Ibata et
al. 1999). Third faint image ``C'' was also reported by Ibata et
al. (1999) between A and B, with a separation of A-C=0\farcs15 and
with the flux ratio of f$_C$/f$_A$=0.18.

Petitjean et al. (2000) suggested a probable damped Lyman-$\alpha$ (DLA)
system at z$_{abs}$=2.974 which has 19.8 $<$ log N(HI) [cm$^{-2}$] $<$
20.3 from data obtained with Keck HIRES (Ellison et al. 1999a,b). The
low metallicity ([X/H] $\sim$ -2.3) derived from weak metal lines (Fe,
Al, Si, C and O) suggest a very small dust amount in the cloud. Thus,
this DLA system is an important object which supports the idea that
there may be a metallicity evolution of DLA in z $>$ 3 (see discussions
in Petitjean et al. 2000, Pettini et al. 1997, 1999).

Another interest of the DLA system is any difference of absorption
features in the line of sight to the images A, B, and C. The small
separation of A-B, which is estimated at $\sim$ 200h$_{75}^{-1}$ pc
(Petitjean et al. 2000), gives us a unique opportunity to investigate
small ($<$ 1 kpc) spatial structure of DLA systems at high-z.  A
systematic study of small-scale structure of high-ionization gas
(\ion{C}{4}) at high redshift was performed by Rauch et al. (2001) by
observing a number of gravitationally lensed QSOs with Keck HIRES. They
found \ion{C}{4} clouds do not seem to have small scale structure for
regions smaller than a few hundred parsecs. Lopez et al. (1999) also
show that \ion{O}{6} is spread out over large scales.

However, there is a strong interest in lower-ionization clouds since
they trace the neutral hydrogen gas (e.g., Wolfe \& Prochaska 2000a)
which should be the source of star-forming molecular clouds. If high-z
DLAs are the progenitor of the present-day disk galaxies (e.g.,
Storrie-Lombardi \& Wolfe 2000 and references therein), the
spatial-structure of low-ionization clouds gives us insights into how
the ``protogalaxies'' are assembled from building blocks (e.g.,
Haehnelt, Steinmetz, \& Rauch 1998, Wolfe \& Prochaska 2000b). Although
Rauch et al. (1999) found a strong variation of low-ionization
\ion{C}{2} and \ion{Si}{2} absorption at z $=$ 3.538 along two
sight-lines separated by only 26h$_{50}^{-1}$ pc for Q1422+231 (z$_{em}
=$ 3.628), it cannot be excluded that the system is somehow associated
with the quasar as discussed by Rauch et al. (1999) because the velocity
difference between the QSO and the system is only 6000 km s$^{-1}$.

\ion{Mg}{2} absorption lines are one of the most promising probes for
studying such low-ionization clouds (Churchill \& Vogt 2001). The
\ion{Mg}{2} $\lambda\lambda$ 2796,2803 doublet absorption lines are most
noted as being associated with galaxies (Bergeron \& Boiss\'{e} 1991;
Steidel, Dickinson \& Persson 1994) and its kinematics for low to
intermediate redshift DLAs are comprehensively studied (Churchill \&
Vogt 2001 and references therein).  However, there has been no detailed
\ion{Mg}{2} absorption line study for high-redshift (z $>$ 2.5) DLAs
because \ion{Mg}{2} doublet absorption line shifts into the
near-infrared ($\lambda >$ 1\,\micron) in the observer's frame and until
recently there has been no capability of high-resolution near-infrared
spectroscopy with a large aperture telescope.

Here we present near-infrared 1.02$-$1.16\,$\mu$m spectra of APM
08279+5255 with 40$-$45 km s$^{-1}$ spectral resolutoin, obtained with
the near-infrared spectrograph IRCS at the Subaru Telescope. In this
rest-frame UV spectra, we have detected \ion{Mg}{2} doublet and
\ion{Fe}{2} absorption lines originated from the DLA system at
z$_{abs}$=2.974. Although future higher resolution ($<$ 10 km s$^{-1}$)
study is in order, the present data gives us a variety of information on
the DLA system with reasonable signal-to-noise.  Also, a difference of
the \ion{Mg}{2} absorption lines between the images A and B was clearly
detected in our spectrum under excellent seeing condition.  In the
following, \S 2 describes the observation and \S 3 describes the results
and preliminary analysis.


\section{Observations}

Near-infrared 1.02$-$1.18\,$\mu$m spectra of APM 08279+5255 were
obtained on 11 January 2001 UT during the scientific commissioning run
of IRCS, a 1$-$5\,$\micron$ near-infrared camera and spectrograph at the
Japanese Subaru 8.2m telescope (Tokunaga et al. 1998, Kobayashi et
al. 2000) on Mauna Kea, Hawaii.  We used the cross-dispersed Echelle
mode of IRCS, which provides a spectral resolution of 15 km s$^{-1}$
($\lambda/\Delta\lambda \sim$ 20,000) with a 0.15-arcsec slit.  The
pixel scale is 0.075-arcsec per pixel in spatial direction and about 7.5
km s$^{-1}$ per pixel in the dispersion direction. The entire IRCS
``zJ-band'' (1.02$-$1.18\,$\mu$m ) was covered simultaneously in one
exposure with the cross-disperser used in order 6 and the echelle in
order 48 to 55. Although seeing was good throughout the observing time
(0\farcs3 to 0\farcs55 in zJ-band), we used a wider slit
(0\farcs6) to maximize the throughput of the slit. Thus, the spectral
resolution was determined by the seeing size and the resultant spectral
resolution was 40 $-$ 45 km s$^{-1}$ ($\lambda/\Delta\lambda \sim$
7,000) in the observed wavelength range.

The slit length was 3.8-arcsec and the slit angle was set to P.A. $=$
32$^\circ$ so that both the lensed images A and B are in the slit. The
object was nodded along the slit by about 2$\arcsec$ after a 300 sec
exposure for sky-background and dark subtraction.  Because of the fine
pixel scale needed for adaptive optics images and the high spectral
resolution, it is impossible to reach the background-limit within a
single exposure (300$-$600 sec) which was less than the time-variation
of the sky OH-emission.  Four sets of data were obtained, resulting in a
total exposure time of 2400 sec.  Spectra of bright telluric standard
stars (HR 4905: A0p, HR 5191: B3V) at similar airmass were obtained in
similar fashion, but only HD 5191 was used in the following analysis
because HD 4905 had several unidentified stellar absorption lines.

The obesrving log is shown in Table 1.  Although we started the
observing right after focusing the secondary mirror of the telescope,
the image quality had gradually changed during the observing due to a
seeing change from 0\farcs30 to 0\farcs55 in zJ-band in an hour. Thus,
we had good image quality only for the first few frames, in which we
could resolve the images A and B with 0\farcs38 separation in the
echellogram (discussed in detail in \S 3.2).

All the data were reduced following standard procedures using the
IRAF\footnote{IRAF is distributed by the National Optical Astronomy
Observatories, which are operated by the Association of Universities for
Research in Astronomy, Inc., under cooperative agreement with the
National Science Foundation} noao.imred.echelle package: sky subtraction
(subtraction of two frames), flat-fielding (halogen lamp with an
integrating sphere), and aperture extraction. Argon lamp spectra which
were taken at the end of the observing night were used for wavelength
calibration. Using atmospheric absorption lines in the object and
standard spectra, we confirmed that the pixel shift along the dispersion
direction was less than 1-pixel throughout the observing time including
the calibration.

\section{Results and Discussions}

\subsection{\ion{Mg}{2} and \ion{Fe}{2} Absorption Lines}

Figure 1 shows 1.02$-$1.16\,$\mu$m spectrum of APM 08279+5255. The raw
spectrum was smoothed to 3 pixels (22.5 km s$^{-1}$) with a boxcar
function for better signal-to-noise.  Strong \ion{Mg}{2}
$\lambda\lambda$2796,2803 absorption lines as well as \ion{Fe}{2}
$\lambda$2587 and $\lambda$2600 absorption lines from the DLA system at
z$_{abs} =$ 2.974 are clearly seen in this spectrum. \ion{Mg}{1}
$\lambda$2853 absorption line is probably detected at the expected
wavelength for the redshift, but qutantitative numbers for this line
should be taken with cautions because of the strong atmospheric
absorption lines at 1.12$-$1.15\,\micron.

The detected absorption lines are summarized in Table 2.  Because the
equivalent width ratio\footnote{The ratio of the \ion{Mg}{2} doublet
equivalent width (W(2796)/W(2803)) should be 2 for non-saturated case
and 1 for fully-saturated case.} W(2796)/W(2803) is about 1.4, the
\ion{Mg}{2} $\lambda$2796 absorption line should have saturated velocity
components. \ion{Fe}{2} $\lambda$2600 may be also saturated judging from
the relatively deep absorption of our low resolution spectrum.
Petitjean et al. (2000) showed that the \ion{Fe}{2} $\lambda$1608 absorption
line has 40\% absorption at the core of the line with $\sim$8 km
s$^{-1}$ resolution.  Because the oscillator strength of \ion{Fe}{2}
$\lambda$2600 (f$=$0.2239; Morton 1991) is about three times larger than
that of \ion{Fe}{2} $\lambda$1608 (f$=$0.064408; Savage \& Sembach 1996), the
\ion{Fe}{2} $\lambda$2600 line is most likely partly saturated.

The column density and metallicity of \ion{Fe}{2}, \ion{Mg}{2}, and
\ion{Mg}{1} estimated from the equivalent widths are also summarized in
Table 2.  In view of the large equivalent widths, \ion{Fe}{2}
$\lambda$2600, \ion{Mg}{2} $\lambda\lambda$2796,2803 lines should be
saturated (see examples in Churchill et al. 1999, Churchill et al. 2000a
and Churchill et al. 2000b). Thus, the estimated column densities for
those lines are a lower-limit. We assumed that \ion{Fe}{2}
$\lambda$2587, \ion{Mg}{1} $\lambda$2583 lines are not saturated. The
following equation for optically thin limit from Savage and Sembach
(1996) was used for estimating the column densities: N[cm$^{-2}$] $=$
1.13 $\times$ 10$^{17}$ W[m\AA] $/$ (f $\lambda^2$[\AA]), where
$\lambda$ is rest wavelength, f is oscillator strength, and W is
equivalent width of the line. The estimated metallicity of \ion{Fe}{2}
and \ion{Mg}{2} in Table 1 is consistent with the result for other metal
lines in Petitjean et al (2000), supporting their conclusion that the
DLA system has quite low-metallicity ([X/H] $\sim$ -2.0 $-$ -2.5) and
may suggest the system's ``young'' nature compared to lower-redshift DLA
systems (Petitjean et al. 2000).

Figure 2 shows the line profile of \ion{Mg}{2} doublet and \ion{Fe}{2}
$\lambda$2600 absorption lines on a velocity scale with the origin at
z$_{abs}$=2.9740.  In our spectrum, all three lines show two major
velocity components at v $\sim$ -10 km s$^{-1}$ (hereafter
``main component'') and $\sim$ +45 km s$^{-1}$ (hereafter
``subcomponent''). Gaussian fit to the two components was performed for
\ion{Mg}{2} doublet and \ion{Fe}{2} $\lambda$2600 using IRAF {\it splot}
task (Table 2). We did not use Voigt profile fitting because the
spectral resolution is not high enough to extract high precision
information. We performed a simple Gaussian fitting with maximum two
components and the decomposition of absorption lines is only
approximate. The fitted Gaussian FWHM suggests that the main component
has a velocity spread no more than 80 km s$^{-1}$. The main component
should be saturated for both \ion{Mg}{2} $\lambda\lambda$ 2796, 2803 and
probably for \ion{Fe}{2} $\lambda$ 2600.  The subcomponent is weaker,
but the equivalent width ratio of the \ion{Mg}{2} doublet is
W$_{sub}$(2796)/W$_{sub}$(2803) $\lesssim$ 1 and this may suggest that
the subcomponent is saturated.  The relatively large equivalent width
(W$_{rest} \sim$0.1\AA\,) also suggests that at least \ion{Mg}{2}
$\lambda$2796 line is saturated (Churchill \& Vogt 2001).

Petitjean et al (2000) identified three major velocity components at
around -28 km s$^{-1}$, +5 km s$^{-1}$, and +45 km s$^{-1}$ for 10 metal
absorption lines in Keck HIRES data with velocity resolution of $\sim$8
km s$^{-1}$. Because Petitjean et al. 2000 does not list the velocity of
the three components, we estimated the velocity from the vertical lines
in their Figure 12. An additional velocity components may be present
between -28 and +5 km s$^{-1}$ components in their Figure 12 and, in
particular, it is clearly seen in Al II $\lambda$1670 absorption
spectrum.  With lower spectral resolution, three velocity components
(-28 km s$^{-1}$, -10 km s$^{-1}$, +5 km s$^{-1}$) should be blended and
the peak position would shift to about -10 km s$^{-1}$. Thus, we
identified our ``main component'' as the mixture of those three
components in Petitjean et al. (2000). Their +45 km s$^{-1}$ component
coincides with our ``subcomponent'' seen in \ion{Mg}{2} $\lambda\lambda$
2796,2803 and \ion{Fe}{2} $\lambda$2600 absorption lines. The
identification is summarized in Table 3.

\subsection{Spatially-resolved Spectra of Images A and B}


Because the seeing was less than 0\farcs4 at the beginning of
observing, two images of the gravitationally lensed QSO, A and B with
0\farcs38 separation, were resolved in the first three frames of 300
sec exposure (see \S 2 and Table 1). In all the three frames it was
noticed that only image A has the ``subcomponent'' at +45 km s$^{-1}$
while both have the ``main component'' in common at around -10 km
s$^{-1}$ (Figure 3). We combined those three frames and extracted
spectra of A and B separately to examine any difference of the DLA
absorption lines. Extracted spectra for A and B (Figure 4) also show the
difference clearly.



Figure 5 shows the cross-cuts of the combined spectrum in spatial
direction. The light grey line shows continuum profile on both sides of
the \ion{Mg}{2} absorption lines while dark grey line shows the profile
at +45 km s$^{-1}$ component. In this plot it is clearly shown that
image A has significantly less flux at +45 km s$^{-1}$ than at the
continuum while image B has almost similar flux as at the continuum. A
slight dip of the dark grey line near image C may suggest that the flux
from image C also sigfinicantly decreased at +45 km s$^{-1}$. However,
that is not clear because the spatial resolution ($\sim$0\farcs35) is
not enough for the small separation between A and C
(0\farcs15). Therefore, we performed a gaussan fitting to the spatial
profiles for \ion{Mg}{2} $\lambda$2796 line (Figure 6). The continuum
profile was fitted well with the combination of three gaussians at A, B,
and C positions with the exact flux ratio as in Ibata et al. (1999). The
profile at +45 km s$^{-1}$ was well fitted similarly with 50\% less flux
for A and 80\% less flux for C (see black lines in the right panel). We
tried to fit the +45 km s$^{-1}$ profile without decreasing the C flux,
but could not reproduce the dip at C-position well (see grey lines in
the same panel). This result suggests that the +45 km s$^{-1}$ component
is present also in image C and significantly decrease the flux from
C. However, in view of the uncertainty of the profile data (see e.g.,
the difference of two continuum profiles for \ion{Mg}{2} 2796 line,
which indicates the uncertainty of the profile in Figure 5), the fitting
results for image C should be treated with caution. In a summary, we
conclude that: (1) image B does not have any absorption from the +45 km
s$^{-1}$ component, (2) image A has $\sim$50\% absorption from the +45
km s$^{-1}$ component, (3) image C may have very strong absorption from
the +45 km s$^{-1}$ component.



Because the estimated distance between images A and B of APM 08279+5255
is only 200h$_{75}^{-1}$ pc at the redshift of the z$=$2.974 DLA system
(Petitjean et al. 2000), our results suggest that there is a small-scale
structure ($<$ 200 pc) in the DLA system. Figure 5 shows that the
fractional change in column density of the subcomponent from A to B is
almost 1, suggesting a sharp cut-off of this subcomponent. If the strong
absorption from the subcomponent truly exists in C as suggested from the
above analysis, it sets a lower limit on the size of the absorbing
cloud of $\sim$80h$_{75}^{-1}$ pc, which corresponds to the separation
of A and C (= 0\farcs15). For high-ion \ion{C}{4} clouds, Rauch et
al. (2001) found that they are mostly featureless on scales below $\sim$
300 pc and the fractional change in column density is mostly less than
0.2 for such small scales (see Figures 8 \& 9 in Rauch et al. 2001). Our
results suggest that low-ionization clouds may have finer structure than
high-ionization clouds at high-z, but a more systematic study is needed
to address this point.


The velocity difference of the subcomponent to the main component is
about 50 km s$^{-1}$ and the subcomponent is classified as ``moderate''
velocity subsystem in the classification of kinematic \ion{Mg}{2}
subsystems by Churchill \& Vogt (2001). The velocity of the subsystem
and the equivalent width is well within the ranges for low to
intermediate redshift DLAs studied by Churchill \& Vogt (2001). They
found that the high or moderate velocity subsystem is probably {\it not}
higher redshift analogues to the Galactic high velocity clouds (HVCs) in
the halo or intergalactic space because the \ion{H}{1} column density of
the substems is too low for HVCs.  Churchill \& Charlton (1999) found
that the \ion{Mg}{2} subsystems are more like compact clouds which are
embedded in extended highly ionized gas like Galactic coronae. From the
ionization modeling of those clouds, they inferred the masses and sizes
of the \ion{Mg}{2} clouds at $\sim$ few $\times$ 10 M$_\sun$ and $\sim$
few $\times$ 10 $-$ 100 pc. Their inferred size is consistent with our
results for the subsystem of the DLA at z$_{abs} = $2.974 ($\lesssim$
200 pc). Although more systematic study is in order, our result presents
the first glimpse of small-scale structure of low-ionization \ion{Mg}{2}
clouds in a DLA system at high redshift.




\acknowledgments

We are grateful to all of the IRCS team members and the Subaru Telescope
observing staff for their efforts that made it possible to obtain these
data. We thank National Astronomical Observatory Japan for the financial
support and encouragement for the construction of the IRCS.




\clearpage


\begin{figure}
\epsscale{0.8}
\plotone{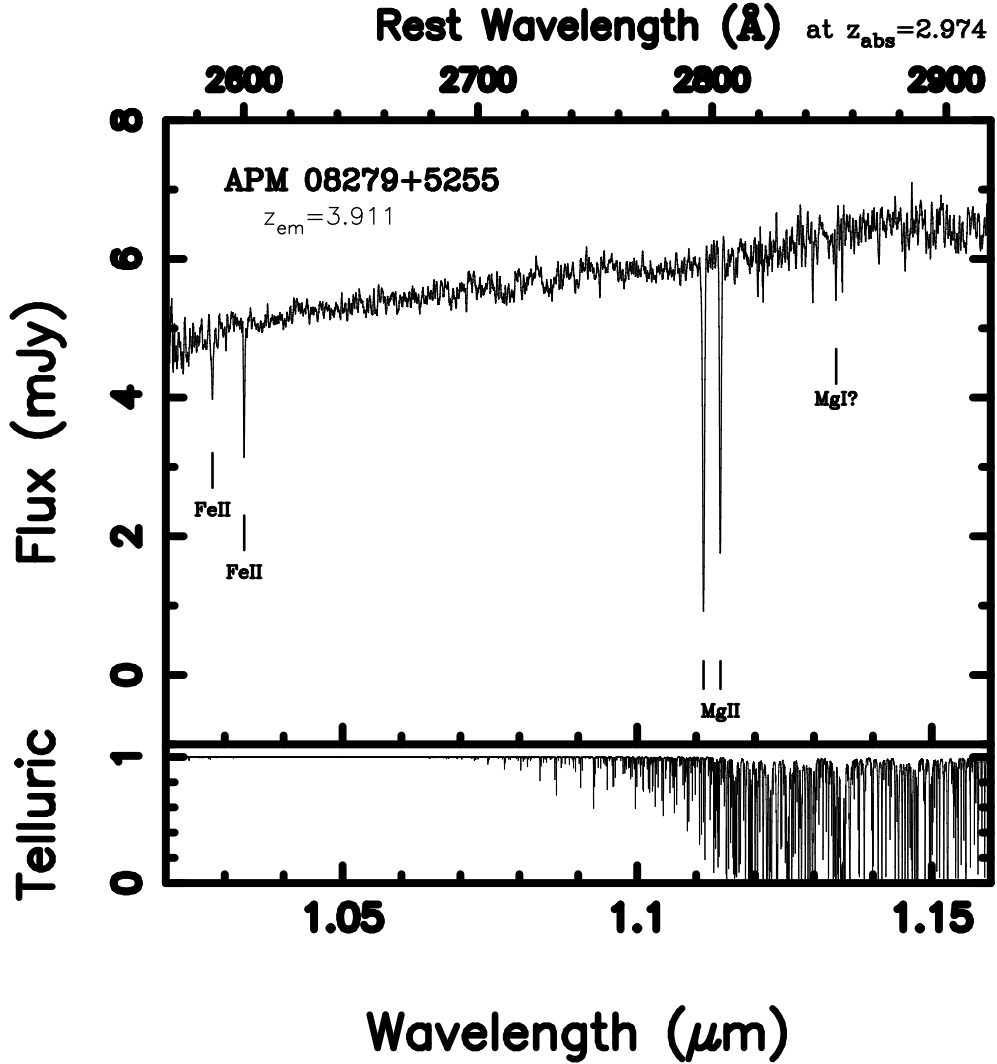} 

\caption{The top figure shows 1.02 $-$ 1.16\,$\mu$m spectrum of APM
08279+5255. All the frames of total integration time of 40 minutes were
used to extract this spectrum. The horizontal axis shows the local-frame
wavelength and the rest-frame wavelength at z$_{abs}=$ 2.974 is shown at
the top.  The spectrum was smoothed with a 3-pixel (22.5 km s$^{-1}$)
boxcar function. \ion{Mg}{2} $\lambda\lambda$2796,2803, \ion{Fe}{2}
$\lambda$2600, \ion{Fe}{2} $\lambda$2587 and probable \ion{Mg}{1}
$\lambda$2852 absorption lines from the DLA system are detected.  The
bottom figure shows the telluric absorption spectrum in the wavelength
region calculated with ATRAN package software (Lord 1992). \label{fig1}}
\end{figure}

\begin{figure}
\epsscale{0.6}
\plotone{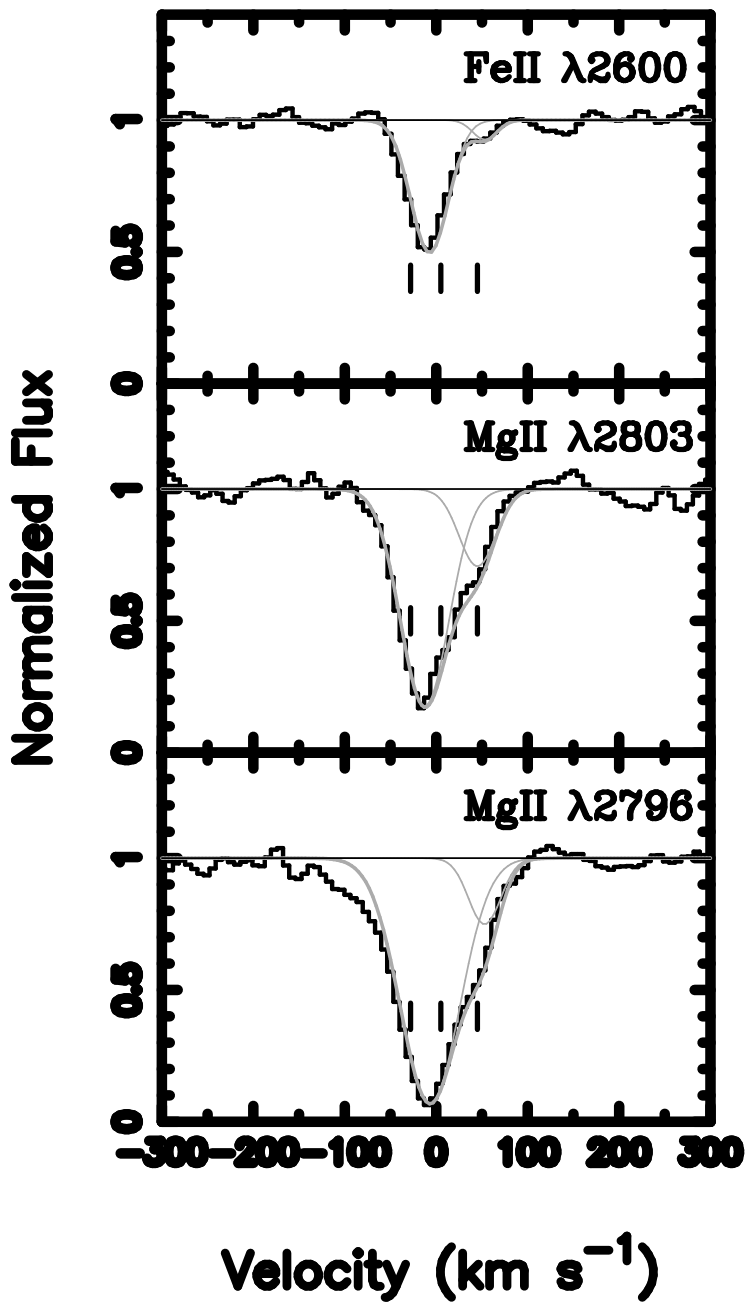} 

\caption{Velocity profile of \ion{Mg}{2} $\lambda\lambda$2796,2803 and
 \ion{Fe}{2} $\lambda$2600 absorption lines from the spectrum shown in
 Figure \ref{fig1}. Velocity resolution is about 40$-$45 km s$^{-1}$
 ($\lambda/\Delta\lambda\sim$7000). The locations of three velocity
 components (-28, +5, +45 km s$^{-1}$) from Petitjean et al. (2000) are
 marked with vertical tick marks. Because of insufficient spectral
 resolution, the -28 km s$^{-1}$ and +5 km s$^{-1}$ components are
 merged and show a single absorption peak at around -10 km s$^{-1}$
 (''main component''). The 45 km s$^{-1}$ component is marginally
 resolved from the strong main component (''subcomponent''). The results
 of two-component Gaussian fitting (see text for detail) are shown with
 grey lines: thin grey lines show each Gaussian component and thick grey
 line shows the sum of the two components. \label{fig2}}
\end{figure}

\begin{figure}
\epsscale{0.6}
\plotone{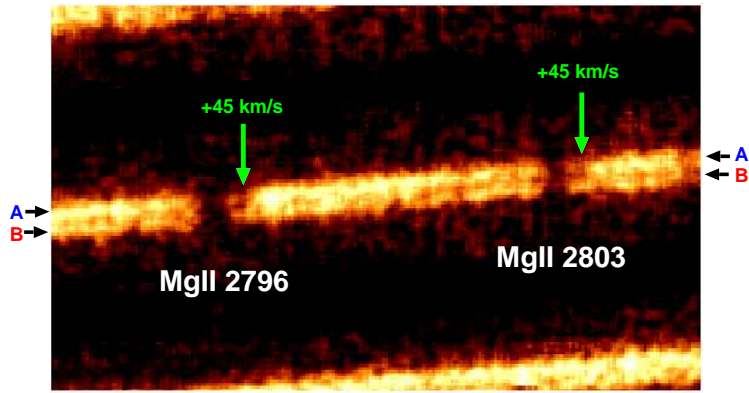}
\notetoeditor{}

 \caption{The echellogram of APM 08279+5255 centered at around
1.1125\,$\mu$m, showing \ion{Mg}{2} doublet absorption lines from the
damped Lyman-$\alpha$ system at z$_{abs}=$ 2.974. The image shown is
from a frame with best seeing (frame No. 1, see Table 1) in which the
QSO images A and B with 0\farcs38 separation were most clearly
resolved. The shown image was sky+dark subtracted (using the second
frame), flat-fielded, then boxcared (3-pix$\sim$0\farcs23) for improving
signal-to-noise.  The absorption subcomponent at +45 km s$^{-1}$ is
marked. This subcomponent is seen in source A, but not in source
B. \label{fig3}}
\end{figure}

\begin{figure}
\epsscale{1.0}
\plotone{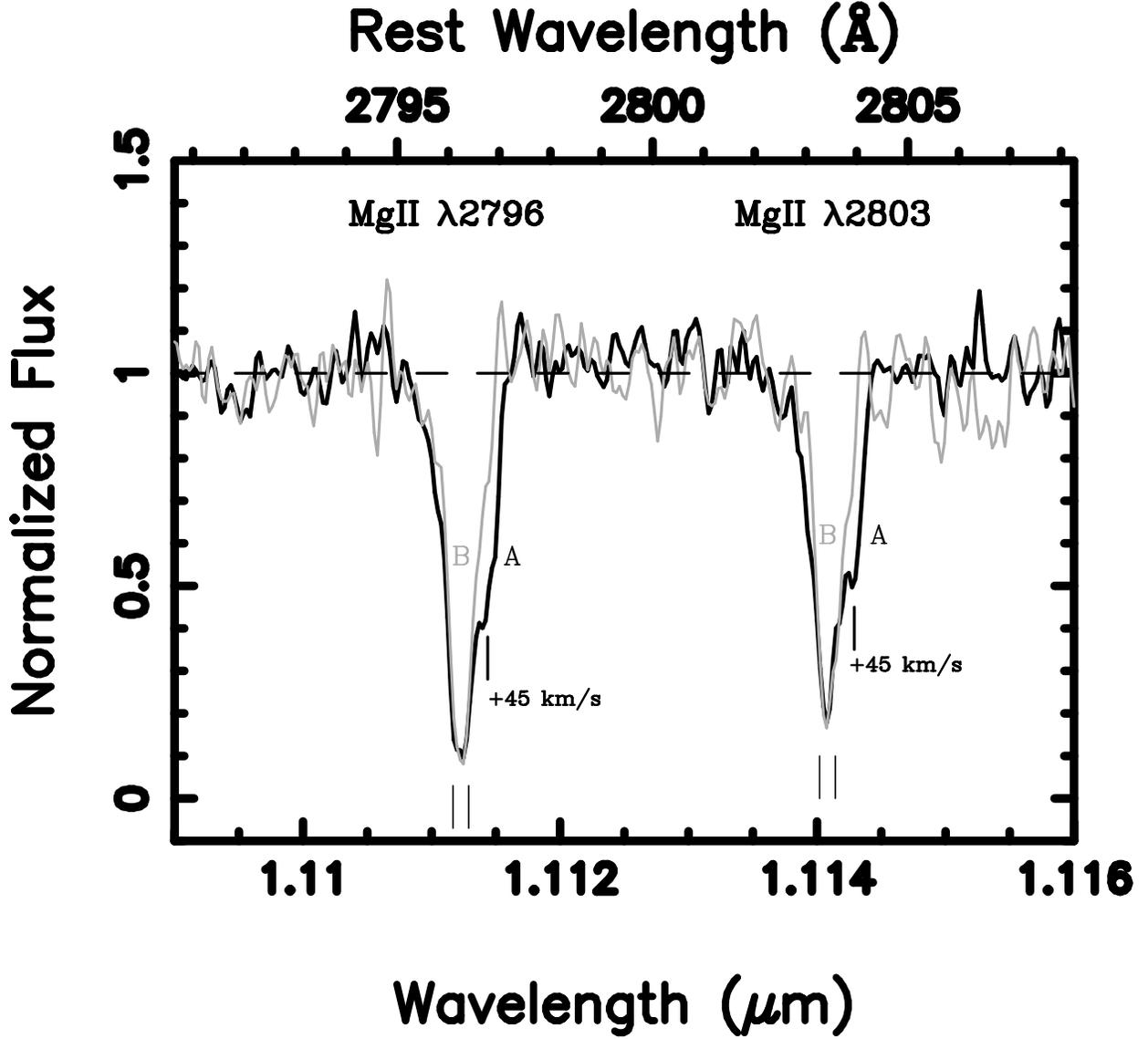}
\notetoeditor{}

\caption{ Extracted spectra of images A and B from a combined
echellogram made from selected frames of good seeing (frames No.1-3, see
Table 1). The horizontal axis shows the local-frame wavelength and the
rest-frame wavelength at z$_{abs}=$ 2.974 is shown at the top. The
vertical axis is normalized flux to the continuum level.  Black and grey
line shows spectrum of source A and B, respectively. Three velocity
components (-28, +5, +45 km s$^{-1}$) from Petitjean et al. (2000) are
marked with thin tick marks.  For both \ion{Mg}{2} $\lambda$2796 and
$\lambda$2803 lines, source A shows the subcomponent at +45 km s$^{-1}$
while B does not. Because of less integration time, signal-to-noise is
not as good as for the spectra in Figure 2. However, the subcomponent at
45 km s$^{-1}$ is clearly detected in source A at more than 10$\sigma$.
\label{fig4}}

\end{figure}

\begin{figure}
\epsscale{0.7}
\plotone{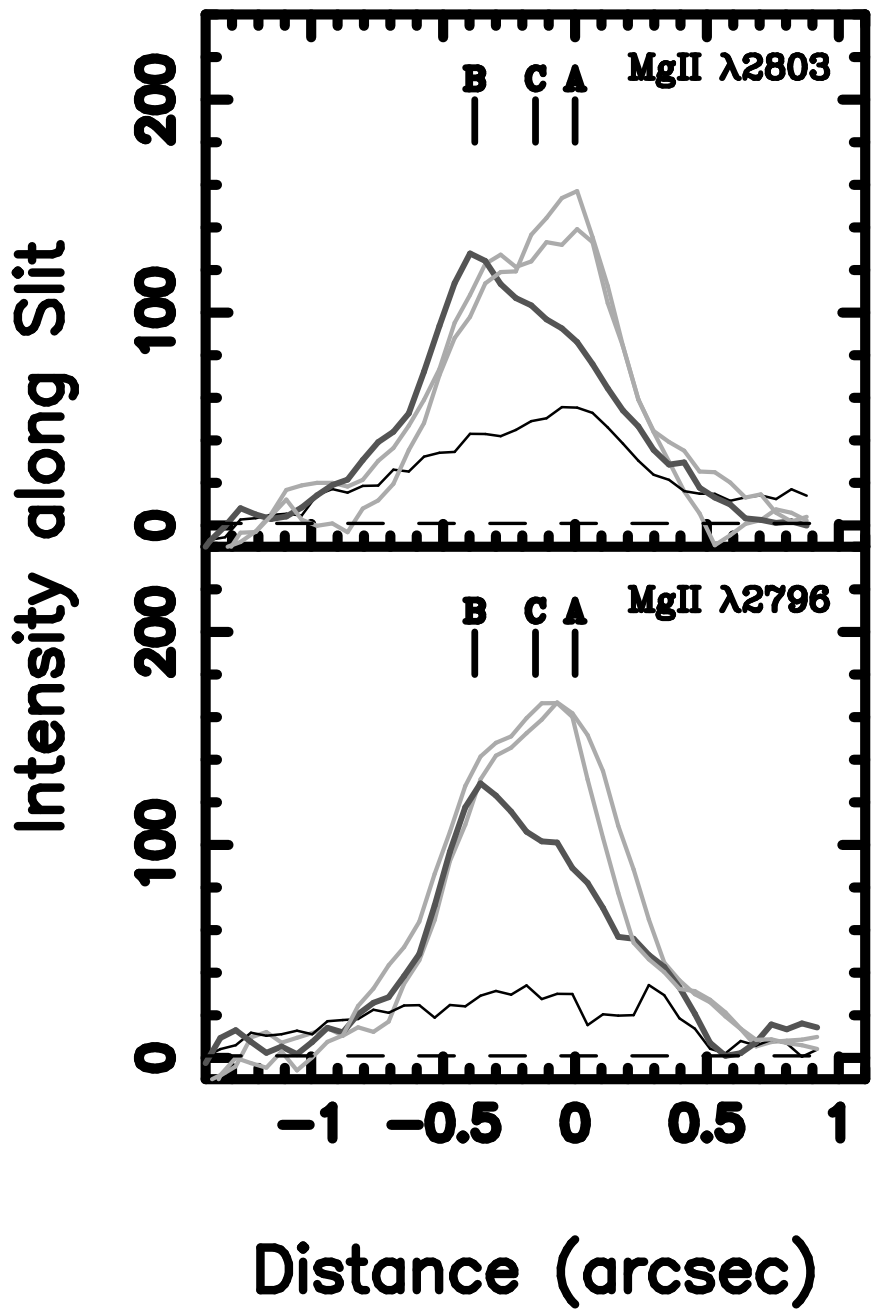}
\notetoeditor{}

\caption{Spatial profile of the \ion{Mg}{2} spectra from a combined
echellogram made from selected frames of good seeing (frames No.1-3, see
Table 1). The horizontal axis shows the distance from the peak of image
A. The vertical axis shows the intensity in arbitrary unit. Dark grey,
thin black, and light grey lines show the profile at the +45 km s$^{-1}$
absorption component, at the bottom of the main absorption component,
and the continuum on both short wavlength side and the long wavelength
side of the absorption lines, respectively. Each profile was made by
binning 4-pixels (= 30 km s$^{-1}$) in the spectrum
direction. Four-pixel boxcar (= 0\farcs3) was applied in the spatial
direction for improving signal-to-noise. The peak intensity of image B
for the 45 km s$^{-1}$ component (dark grey line) does not change from
that for the continuum (light grey line). However, the peak intensity of
image A for the 45 km s$^{-1}$ component decreases significantly from
the continuum level.  \label{fig5}}

\end{figure}

\begin{figure}
\epsscale{1.0}
\plotone{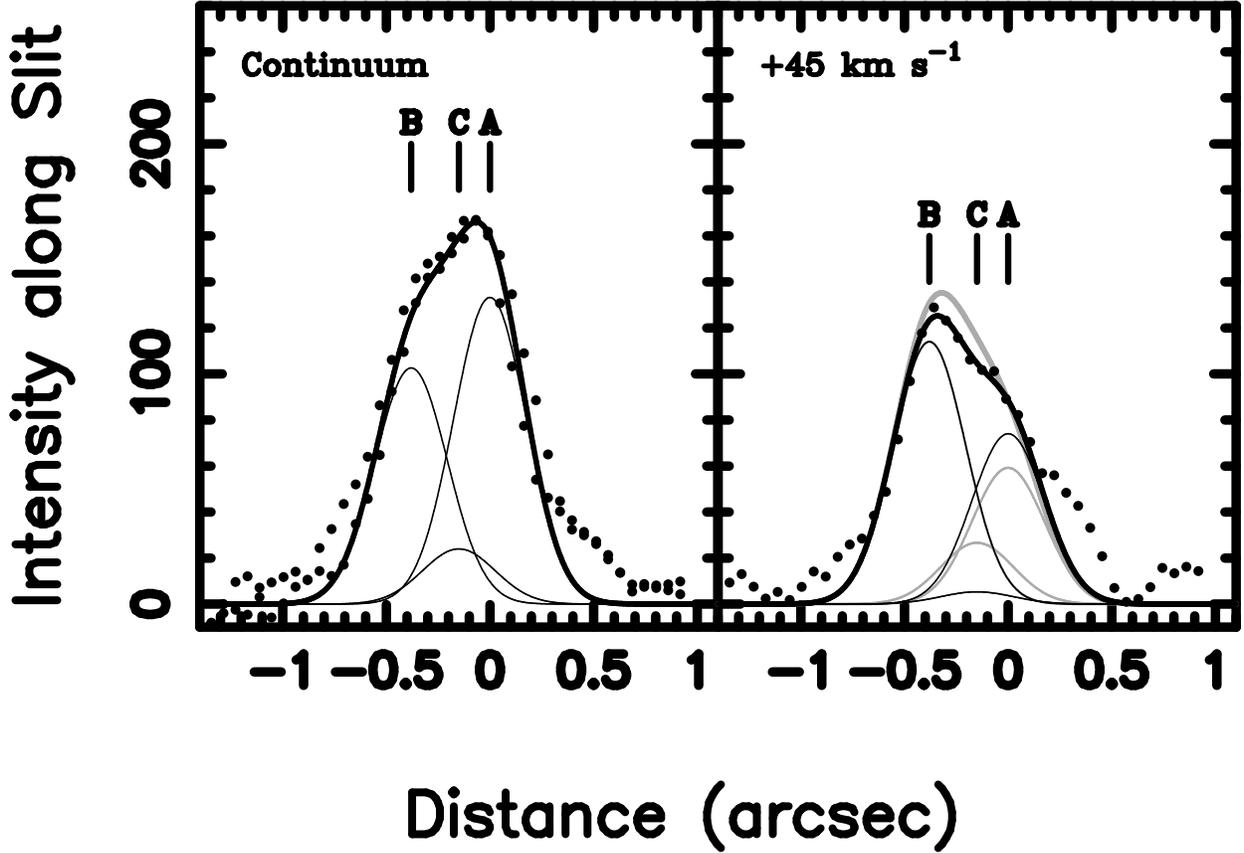}
\notetoeditor{}

\caption{Three component gaussian fit to the spatial profiles in Figure
5. Black dots show the data points and the thick line shows the model
fit. Thin lines show each gaussian component with peak at A, B, and C
positions. The full-width-half-maximum of the gaussians is 0\farcs35.
The continuum profile (left) can be well fitted with three gaussian
components with the flux ratio of the three images (f$_A$ : f$_B$ :
f$_C$ = 1.30 : 1.00 : 0.23) as in Ibata et al (1999). The profile at +45
km s$^{-1}$ (right) was well fitted similarly with three gaussians with
f$_A$ : f$_B$ : f$_C$ = 0.65 : 1.00 : 0.05 (see black lines in the right
panel). We tried to fit the +45 km s$^{-1}$ profile with the same C flux
as in the continuum profile, but could not reproduce the ``dip'' of the
prifile at C-position well (grey lines show the fit with f$_A$ : f$_B$ :
f$_C$ = 0.52 : 1.00 : 0.23).  \label{fig6}}

\end{figure}

\clearpage 






\clearpage

\begin{deluxetable}{ccccccccc}
\tabletypesize{\scriptsize}
\tablecaption{Observing Log for APM 08279+5255
\label{tbl-1}}
\tablewidth{0pt}
\tablehead{
\colhead{Frame No.} &
\colhead{Dither\tablenotemark{a}} &
\colhead{Time\tablenotemark{b}} &
\colhead{Integration time} &
\colhead{Seeing\tablenotemark{c}} &
\colhead{Selection\tablenotemark{d}}\\
\colhead{} &
\colhead{} &
\colhead{} &
\colhead{(sec)} &
\colhead{(arcsec)} &
\colhead{} &
\colhead{}
}
\startdata
1 & a & 13:00       & 300 & 0.30 & Yes\\
2 & b & 13:06       & 300 & 0.40 & Yes\\
3 & b & 13:11       & 300 & 0.35 & Yes\\
4 & a & 13:17       & 300 & 0.50 & --\\
5 & a & 13:44       & 300 & 0.55 & --\\
6 & b & 13:50       & 300 & 0.55 & --\\
7 & b & 13:55       & 300 & 0.55 & --\\
8 & a & 14:01       & 300 & 0.55 & --\\
\enddata


\tablenotetext{a}{Dithering positions of the object on the
 slit. Positions ``a'' and ``b'' have roughly 2$\arcsec$ offset. Pairs
 of ``a'' and ``b'' was used for sky subtraction.}

\tablenotetext{b}{UT time for exposure start. All data were obtained on
2001 Jan. 11 UT.}

\tablenotetext{c}{Seeing in zJ band. Estimated from the FWHM of the
 spectra in the slit direction. Because the FWHM was estimated from
 profile fitting of A plus B assuming the separation of 0\farcs38,
 the accuracy is not better than 1 pix ($\sim$0\farcs08).
}

\tablenotetext{d}{Selected frames with good seeing. Frame No.1 was used
 for making Figure 3. Frames No.1-3 were used for making Figures 4, 5,
 and 6.}

\end{deluxetable}

\begin{deluxetable}{ccccccccc}
\tabletypesize{\scriptsize}
\tablecaption{Absorption Lines in the System at z$_{abs}=$ 2.974. 
\label{tbl-2}}
\tablewidth{0pt}
\tablehead{
\colhead{Species} &
\colhead{Line} &
\colhead{W$_{obs}$\tablenotemark{a}} &
\colhead{W$_{rest}$\tablenotemark{b}} &
\colhead{f\tablenotemark{c}} &
\colhead{Column Density\tablenotemark{d}} &
\colhead{log(X/H)\tablenotemark{e}} &
\colhead{log(X/H)$_\odot$\tablenotemark{f}} &
\colhead{[X/H]\tablenotemark{g}}\\
\colhead{} &
\colhead{} &
\colhead{(\AA)} &
\colhead{(\AA)} &
\colhead{} &
\colhead{(cm$^{-2}$)} &
\colhead{} &
\colhead{} &
\colhead{}
}
\startdata
Fe II & $\lambda$2587 & 0.53 & 0.13 & 0.0684 & \,\,\,\,13.5 & \,\,\,\,-6.8 & -4.49 & \,\,\,\,-2.3\\
Fe II & $\lambda$2600 & 0.97 & 0.24 & 0.2239 & $>$13.3 & $>$-7.0 & -4.49 & $>$-2.6\\
Mg II & $\lambda$2796 & 3.44 & 0.87 & 0.6123 & $>$13.3 & $>$-7.0 & -4.42 & $>$-2.6\\
Mg II & $\lambda$2803 & 2.46 & 0.62 & 0.3054 & $>$13.5 & $>$-6.8 & -4.42 & $>$-2.4\\
Mg I  & $\lambda$2853 & 0.29 & 0.07 & 1.81 & \,\,\,\,11.7 & \,\,\,\,-8.6 & -4.42 & \,\,\,\,-4.2\\
 \enddata


\tablenotetext{a}{W$_{obs}$: Total equivalent width in the observer
 frame. Typical uncertainty is 0.07\AA.}
\tablenotetext{b}{W$_{rest}$: Total equivalent width in the rest frame (=
W$_{obs}$/(1+z$_{abs}$) ). Typical uncertainty is 0.02\AA.}
\tablenotetext{c}{Oscillator strength from Morton (1991) except for Fe
 II $\lambda$2587 which is from Savage \& Sembach (1996)}
\tablenotetext{d}{See text for detail.}
\tablenotetext{e}{Assuming log N(H) [cm$^{-2}$] $=$ log N(H I) [cm$^{-2}$] $=$ 20.3  (Petitjean et al. 2000)}
\tablenotetext{f}{From Savage \& Sembach (1996)} 
\tablenotetext{g}{[X/H] $=$ log(X/H) $-$ log(X/H)$_\odot$
}
\end{deluxetable}



%
%

\begin{deluxetable}{rccccccccccc}
\tabletypesize{\scriptsize}
\tablecaption{Velocity Components in Three Absorption Lines 
\label{tbl-3}}
\tablewidth{0pt}
\tablehead{
\colhead{Line} &
\colhead{Rest Wavelength\tablenotemark{a}} &
\colhead{z} &
\colhead{FWHM\tablenotemark{b}} &
\colhead{W$_{obs}$\tablenotemark{c}} &
\colhead{ID\tablenotemark{d}}\\
\colhead{} &
\colhead{(\AA)} &
\colhead{} &
\colhead{(km s$^{-1}$)} &
\colhead{(\AA)} &
\colhead{}
}
\startdata
Fe II $\lambda$2600 & 2600.173 & 2.973909 $\pm$ 0.000012
& 48 $\pm$ 3 & 0.89 $\pm$ 0.04
& main\\
                    &          & 2.974709 $\pm$ 0.000093
& 31 $\pm$ 15 & 0.09 $\pm$ 0.03
& sub\\
\tableline
Mg II $\lambda$2796 & 2796.352 & 2.973913 $\pm$ 0.000012
& 76 $\pm$ 3 & 2.81 $\pm$ 0.08
& main\\
                    &          & 2.974703 $\pm$ 0.000030
& 41 $\pm$ 5 & 0.41 $\pm$ 0.07
& sub\\
\tableline
Mg II $\lambda$2803 & 2803.531 & 2.973835 $\pm$ 0.000015
& 62 $\pm$ 2 & 2.00 $\pm$ 0.07
& main\\
                    &          & 2.974598 $\pm$ 0.000037
& 47 $\pm$ 5 & 0.54 $\pm$ 0.07
& sub\\
\tableline
 \enddata


\tablenotetext{a}{From Morton (1991)}

\tablenotetext{b}{Gaussian FWHM of the absorption line}

\tablenotetext{c}{W$_{obs}$: Equivalent width in the observer frame}

\tablenotetext{d}{Identification with the velocity components in
 Petitjean et al. (2000). ``Main Component'' is a combination of the
 systems spanning from -28 km s$^{-1}$ to +5 km s$^{-1}$ which cannot be
 resolved with the spectral resolution of the present
 data. ``Subcomponent'' is a subsystem at v $=$ +45 km s$^{-1}$.}


\end{deluxetable}

\end{document}